\documentclass[12pt]{iopart}
\pdfoutput=1
\usepackage{graphicx}

\begin{document}

\title[The Large-Volume Limit of a Quantum Tetrahedron]{The Large-Volume Limit of a Quantum Tetrahedron is a Quantum Harmonic Oscillator}

\author{John Schliemann}

\address{Institute for Theoretical Physics, University of Regensburg,
D-93040 Regensburg, Germany}
\ead{john.schliemann@physik.uni-regensburg.de}
\begin{abstract}
It is shown that the volume operator of a quantum tetrahedron is, in the
sector of large eigenvalues, accurately described by a quantum harmonic
oscillator. This result relies on the fact that (i) the volume operator couples
only neighboring states of its standard basis, and (ii) its matrix elements
show a unique maximum as a function of internal angular momentum
quantum numbers. These quantum numbers, considered as a continuous variable,
are the coordinate of the oscillator describing its quadratic potential, 
while the corresponding
derivative defines a momentum operator. We also analyze the scaling
properties of the oscillator parameters as a function of the size
of the tetrahedron, and the role of different angular momentum coupling
schemes.
\end{abstract}

\section{Introduction}

The quantum volume operator is one of the most studied objects in the field
of loop quantum gravity and of crucial importance for the construction
of dynamics within this approach \cite{Rovelli04,Thiemann07,Perez13}. 
In the literature, one finds traditionally two versions
of such an operator, due to Rovelli and Smolin \cite{Rovelli95},
and to Ashtekar and Lewandowski \cite{Ashtekar95}, respectively.
Their properties and interrelations have been intensively
investigated \cite{Loll95,Loll96,DePietri96,DePietri97,Thiemann98,Barbieri98,Carbone02,Brunnemann06,Giesel06a,Giesel06b,Meissner06,Brunnemann08a,Brunnemann08b,Dittrich09,Bianchi11a,Bianchi11b,Bianchi12,Aquilanti13},
including a third proposal for a volume operator by Bianchi, Dona, and 
Speziale \cite{Bianchi11a}. The latter one 
is closer to the concept of spin foams \cite{Perez13} and relies on
an older geometric theorem due to Minkowski \cite{Minkowski1897}.  
Volume operators are usually considered in connection with
polyhedra. The most elementary objects of this kind are tetrahedra consisting
of four faces which are represented by angular momentum operators coupling
to a total spin singlet \cite{Barbieri98,Baez99}. 
Here all three definitions of the volume
operator coincide. Among the most recent developments, Bianchi and
Haggard have performed a Bohr-Sommerfeld quantization of the volume
using an appropriate parameterization of the classical phase space
of a tetrahedron, and the obtained semiclassical eigenvalues agree
amazingly well with exact numerical data \cite{Bianchi11b,Bianchi12}. 

The purpose of the present communication is to point out that, in
the sector of large eigenvalues, the volume operator of such a quantum
tetrahedron is accurately described by a quantum harmonic oscillator.
Our presentation will continue as follows: After briefly summarizing 
important features of the quantum tetrahedron and its volume operator
in section \ref{tet}, we derive our central result, starting from 
numerical observations, in section \ref{vol}. We give explicit formulae
for the large-eigenvalue sector of the (square of the) volume operator
and also analyze its scaling behavior as a function of the tetrahedron size.
In section \ref{rec} we discuss the role of different angular momentum
coupling schemes, and in section \ref{concl} we close with an outlook.

\section{The Quantum Tetrahedron}
\label{tet}

A quantum tetrahedron consists of four angular momenta $\vec j_i$,
$i\in\{1,2,3,4\}$ representing its faces and coupling to a vanishing total
angular momentum \cite{Barbieri98,Carbone02,Baez99,Bianchi11b,Bianchi12}
, i.e. the Hilbert space consists of all states $|k\rangle$
fulfilling
\begin{equation}
\left(\vec j_1+\vec j_2+\vec j_3+\vec j_4\right)|k\rangle=0\,.
\end{equation}
In what follows we will adopt the coupling scheme where both pairs
$\vec j_1,\vec j_2$ and $\vec j_3,\vec j_4$ couple first
to two irreducible SU(2) representations of dimension $2k+1$ each,
which are then added to give a singlet. Thus, the quantum number $k$
ranges as $k_{\rm min}\leq k\leq k_{\rm max}$ with
\begin{equation}
k_{\rm min}=\max\{|j_1-j_2|,|j_3-j_4|\}\quad,\quad k_{\rm max}=\min\{j_1+j_2,j_3+j_4\}
\,,
\end{equation}
leading to a total dimension of $d=k_{\rm max}-k_{\rm min}+1$.
The volume operator can be formulated as
\begin{equation}
V=\frac{\sqrt{2}}{3}\sqrt{|\vec E_1\cdot(\vec E_2\times\vec E_3)|}
\label{defV}
\end{equation}
where the operators
\begin{equation}
\vec E_i=8\pi\gamma\ell_P^2\vec j_i\,,
\end{equation}
$i\in\{1,2,3,4\}$ represent the faces of the tetrahedron with
$\ell_P^2=\hbar G/c^3$ and $\gamma$ being the Immirzi parameter.
As seen form Eq.~(\ref{defV}) it is useful to consider the operator
\begin{equation}
\tilde Q=\vec E_1\cdot(\vec E_2\times\vec E_3)
\end{equation}
which, in the basis of the states $|k\rangle$, can be represented as
\cite{Carbone02,Bianchi12,Levy-Leblond65,Chakrabarti64,Edmonds57}
\begin{equation}
\tilde Q=\sum_{k=k_{\rm min}+1}^{k_{\rm max}}i\alpha(k)
\left(|k\rangle\langle k-1|-|k-1\rangle\langle k|\right)
\end{equation}
with
\begin{equation}
\alpha(k)=2\frac{\Delta(k,j_1+1/2,j_2+1/2)\Delta(k,j_3+1/2,j_4+1/2)}
{\sqrt{k^2-1/4}}\,.
\label{alpha}
\end{equation}
Here $\Delta(a,b,c)$ is the area of a triangle with edges $a,b,c$ expressed
via Heron\rq s formula,
\begin{equation}
\Delta(a,b,c)=\frac{1}{4}\sqrt{(a+b+c)(-a+b+c)(a-b+c)(a+b-c)}\,.
\end{equation}
Note that $\tilde Q$ couples only basis states $|k\rangle$ with neighboring
labels.
In the following it will be convenient to readjust the phases of these
states via the unitary matrix
$u_{\pm}={\rm diag}(1,\pm i,-1,\mp i,1,\dots)$ such that the resulting
operator becomes real,
\begin{equation}
u_{\pm}\tilde Qu_{\pm}^+=:\mp Q=\mp\sum_{k=k_{\rm min}+1}^{k_{\rm max}}\alpha(k)
\left(|k\rangle\langle k-1|+|k-1\rangle\langle k|\right)\,.
\label{defQ}
\end{equation}
Since $\tilde Q$ is antisymmetric, the spectrum of $\tilde Q$ and, in turn, $Q$
consists for even $d$ of pairs of eigenvalues $q,(-q)$ differing in sign.
Moreover, because of
\begin{equation}
u\tilde Qu^+=-\tilde Q\quad,\quad uQu^+=-Q
\end{equation} 
with $u=(u_{\pm})^2={\rm diag}(1,-1,1,\dots)$, the corresponding eigenstates
$|\phi_q\rangle$, $|\phi_{-q}\rangle$ fulfill
\begin{equation}
|\phi_{-q}\rangle=u|\phi_q\rangle\,.
\label{signchange}
\end{equation} 
For odd $d$ an additional zero eigenvalue occurs whose eigenvector
(with respect to $\tilde Q$) has the unnormalized form \cite{Brunnemann06}
\begin{eqnarray}
|\phi_0\rangle & \propto & 
\Biggl(1,0,\frac{\alpha(k_{\rm min}+1)}{\alpha(k_{\rm min}+2)},0,
\frac{\alpha(k_{\rm min}+1)\alpha(k_{\rm min}+3)}
{\alpha(k_{\rm min}+2)\alpha(k_{\rm min}+4)},0\nonumber\\
 & & \quad\dots,
\frac{\alpha(k_{\rm min}+1)\alpha(k_{\rm min}+3)\cdots\alpha(k_{\rm max}-1)}
{\alpha(k_{\rm min}+2)\alpha(k_{\rm min}+4)\cdots\alpha(k_{\rm max})}
\Biggr)\,,
\end{eqnarray}
which is, as it must be, an eigenstate of $u$.

\section{Large-Volume Limit}
\label{vol}
\begin{figure}
 \includegraphics[width=\columnwidth]{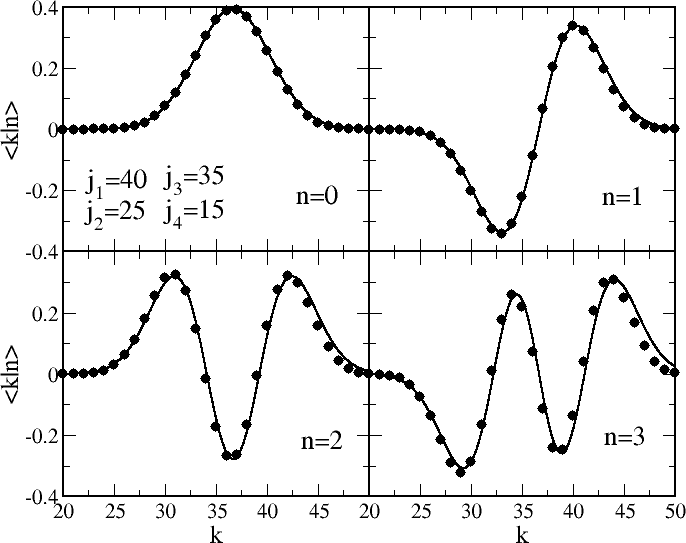}
\caption{The coefficients $\langle k|n\rangle$ (filled circles)
for small $n$ and a typical choice of angular momentum quantum numbers.
The solid lines are oscillator wave functions $\psi_n(k-\bar k+1/2;\omega)$
according to Eq.~(\ref{oscwf}).}
\label{fig1}
\end{figure}
\begin{figure}
 \includegraphics[width=\columnwidth]{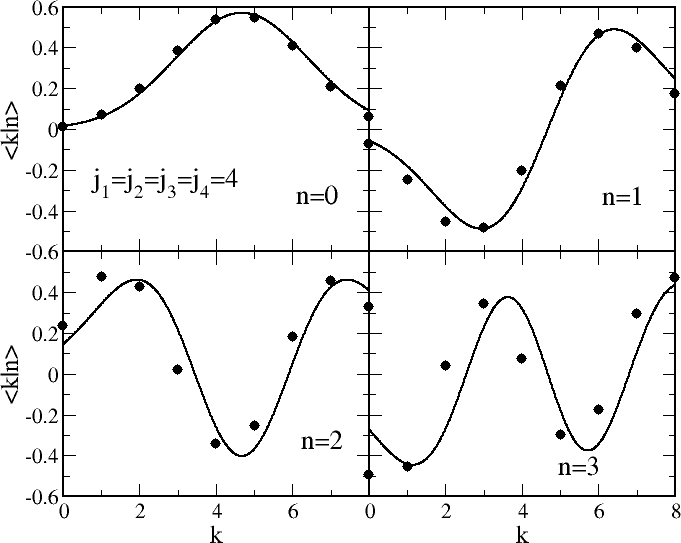}
\caption{The coefficients $\langle k|n\rangle$ (filled circles)
for small $n$ and $j_i\equiv 4$.
The solid lines are oscillator wave functions $\psi_n(k-\bar k+1/2;\omega)$
according to Eq.~(\ref{oscwf}).}
\label{fig2}
\end{figure}
We denote by $|n\rangle$, $n\in\{0,1,2,\dots\}$, the eigenstates of $Q$ 
in descending order of eigenvalues with $|0\rangle$ being the state
with the largest eigenvalue. In the above basis they can
be expanded as 
\begin{equation}
|n\rangle=\sum_{k=k_{\rm min}}^{k_{\rm max}}\langle k|n\rangle|k\rangle
\end{equation}
where the coefficients $\langle k|n\rangle$ can be viewed as the 
``wave function'' of the state $|n\rangle$ with respect to the
``coordinate'' $k$. Fig.~\ref{fig1} shows this data for small $n$ and
a typical choice of angular momentum quantum numbers (all being of order
a few ten). As seen there, the functions $\langle k|n\rangle$ show the
characteristic features of wave functions of the harmonic oscillator for
low-lying states. Indeed, the solid lines in Fig.~\ref{fig1} are
gauss-hermitian oscillator wave functions for parameters to be determined
a few lines below. Such properties of the functions
$\langle k|n\rangle$ occur for arbitrary sufficiently large angular
momentum quantum numbers $j_i$ and sets in when all $j_i$ exceed a value
of about five. For illustration, Fig.~\ref{fig2} displays the data for the
case $j_i\equiv 4$ where the oscillator-like features of the wave functions
gradually disappear with increasing $n$.

\begin{figure}
 \includegraphics[width=\columnwidth]{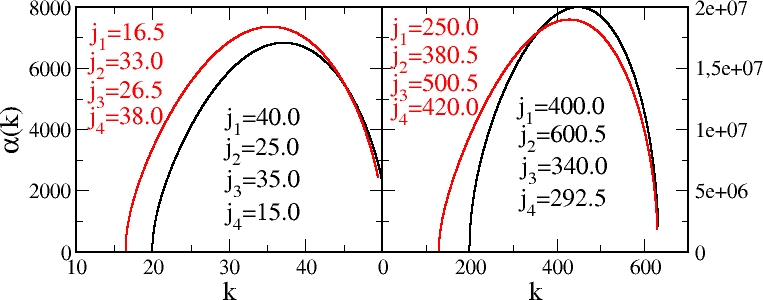}
\caption{The matrix elements $\alpha(k)$ for various choices of angular 
momentum quantum numbers. The left panel includes the situation of 
Fig.~\ref{fig1}. In all cases minima occur at $k\in\{k_{\rm min}+1,k_{\rm max}\}$
with a unique maximum in between.}
\label{fig3}
\end{figure}

The observation made in Figs.~\ref{fig1} and \ref{fig2} can be explained as 
follows: Fig.~\ref{fig3} shows the matrix elements $\alpha(k)$ as a function
of $k$ for several arbitrary choices of angular momentum lengths including
the situation of Fig.~\ref{fig1}. In all cases, 
minima occur at $k\in\{k_{\rm min}+1,k_{\rm max}\}$ with a unique maximum in between
at $k=\bar k$ determined by
\begin{equation}
\left(\frac{d\alpha(k)}{dk}\right)_{k=\bar k}=0\,,
\end {equation}
where we have considered $k$ as a continuous variable. 
The above features can also be established by a detailed 
analytical discussion of the function $\alpha(k)$. 

Now, since the operator
$Q$ couples 
only states with neighboring label $k$, the 
wave functions with large eigenvalues will have
predominantly support around the maximum of $\alpha(k)$. We therefore
expand the matrix elements of $Q$ between arbitrary states 
$|\Phi\rangle$, $|\Psi\rangle$
(lying predominantly in the sector of large eigenvalues) around $\bar k$,
i.e. $k=\bar k-1/2+x$, where the decrement of $(1/2)$ accounts for the
fact that $\alpha(\bar k)$ couples states of the form $|\bar k-1\rangle$
and $|\bar k\rangle$. In doing so, we obtain  
\begin{eqnarray}
 & & \langle\Phi|Q|\Psi\rangle 
=\sum_k\alpha(k)\left(\langle\Phi|k\rangle\langle k-1|\Psi\rangle
+\langle\Phi|k-1\rangle\langle k|\Psi\rangle\right)\nonumber\\
 & & \qquad\approx\int dx\,\Phi^*(x)
\left(2\alpha(\bar k)+\alpha(\bar k)\frac{d^2}{dx^2}
+\left(\frac{d^2\alpha(k)}{dk^2}\right)_{k=\bar k}x^2\right)\Psi(x)\,.
\label{Qmat}
\end{eqnarray}
Here we have introduced the notations $\Phi(x)=\langle\bar k+x|\Phi\rangle$,
$\Psi(x)=\langle\bar k+x|\Psi\rangle$, and additionally performed
a continuum approximation to the latter function
according to
\begin{equation}
\langle k+1|\Psi\rangle+\langle k-1|\Psi\rangle-
2\langle k|\Psi\rangle\approx
\frac{d^2\Psi(x)}{dx^2}\,.
\end {equation}
From Eq.~(\ref{Qmat}) one easily reads off an effective
operator having the form of a harmonic oscillator,
\begin{equation}
Q_{\rm osc}=\bar q\left[1-\left(-\frac{1}{2}\frac{d^2}{dx^2}
+\frac{\omega^2}{2}x^2\right)\right]
\label{Qeff}
\end{equation}
with
\begin{equation}
\bar q=2\alpha(\bar k)
\label{barq}
\end{equation}
and 
\begin{equation}
\omega^2=-\frac{\left(\frac{d^2\alpha(k)}{dk^2}\right)_{k=\bar k}}
{\alpha(\bar k)}>0\,.
\label{omega}
\end{equation}
The eigenstates $\psi_n(x)=\langle x|n\rangle$ 
of $Q_{\rm eff}$ are just the well-known wave functions
diagonalizing the harmonic oscillator in real-space representation,
\begin{equation}
\psi_n(x;\omega)
=\sqrt{\frac{1}{n!2^n}\sqrt{\frac{\omega}{\pi}}}H_n(\sqrt{\omega}x)
e^{-\omega x^2/2}
\label{oscwf}
\end{equation}
where $H_n(x)$ are the usual Hermite polynomials. These functions 
$\psi_n(x;\omega)=\psi_n(k-\bar k+1/2;\omega)$ are plotted as
solid lines in Fig.~\ref{fig1} and are remarkably accurate approximations
to the coefficients $\langle k|n\rangle$.
The corresponding
eigenvalues are
\begin{equation}
q_n^{\rm osc}=\bar q\left(1-\omega(n+1/2)\right)\,.
\label{oscev}
\end{equation}
\begin{table}
\begin{center}
\begin{tabular}{c|c|c|c|}
 n & $q_n$ & $q_n^{\rm osc}$ & $(q_n-q_n^{\rm osc})/q_n$ \\
\hline
 0 & 13141.3 & 13136.3 & $3.8\cdot 10^{-4}$ \\
 1 & 12135.3 & 12109.8 & $2.1\cdot 10^{-3}$ \\
 2 & 11149.4 & 11083.3 & $5.9\cdot 10^{-3}$ \\
 3 & 10183.6 & 10056.7 & $1.2\cdot 10^{-2}$ 
\end{tabular}
\end{center}
\vspace{0.2cm}
\caption{The largest eigenvalues $q_n$ of $Q$ obtained by numerical 
diagonalization of the operator, and the corresponding approximate
eigenvalues $q_n^{\rm osc}$ according to Eq.~(\ref{oscev}).
The choice of angular momentum quantum numbers is the same as in 
Fig.~\ref{fig1}.
The exact and the approximate data agree within a few per mille.
\label{table1}}
\end{table}
Table~\ref{table1} compares the largest eigenvalues of $Q$ obtained via
exact numerical diagonalization with the approximate results Eq.~(\ref{oscev}).
Both data coincide within a few per mille, in accordance with our previous
findings regarding the corresponding eigenvectors. Note that under a rescaling
of all four angular momenta, $j_i\mapsto uj_i$, $\bar q$ scales in leading
order as $u^3$, while for the frequency one finds $\omega\propto 1/u$.

In summary, we have constructed an effective operator describing the sector 
of large eigenvalues of the square of the volume operator of a quantum
tetrahedron. This operator has the form of a harmonic
oscillator with a \lq coordinate\rq  $x$ and a \lq momentum\rq
\begin{equation}
p=-i\frac{d}{dx}
\end{equation}
fulfilling the commutation relation
\begin{equation}
[p,x]=-i
\end{equation}
which is part of the bedrock of quantum theory.

The approximate data shown in Fig.~\ref{fig1} (solid lines)
and table~\ref{table1} was generated by first finding numerically the
maximum position $\bar k$ of $\alpha(k)$ and inserting this value into
an analytical expression of $(d^2\alpha/dk^2)$ to obtain $\omega$ via
Eq.~(\ref{omega}). Thus, no adjustable parameter is involved.
Closed analytical results for $\bar k$ are possible if the four
angular momenta come in two pairs of equal length, and the expressions
become particularly simple in the case of a regular tetrahedron,
$j_1=j_2=j_3=j_4=:j$. Here one has
\begin{eqnarray}
\bar k^2 & = & \frac{2}{3}j(j+1)+\frac{1}{3}
+\frac{2}{3}\sqrt{(j(j+1))^2-\frac{1}{2}j(j+1)-\frac{1}{8}}\\
& = & \frac{4}{3}j(j+1)+\frac{1}{6}+{\cal O}\left(\frac{1}{j}\right)
\end{eqnarray}
such that the parameters entering the effective operator (\ref{Qeff})
are given, to leading orders in $j$, by
\begin{equation}
\bar q=\frac{4}{3\sqrt{3}}\left(j(j+1)\right)^{3/2}
+{\cal O}\left(j\right)\,,
\end{equation}
\begin{equation}
\left(\frac{d^2\alpha(k)}{dk^2}\right)_{k=\bar k}=
-\sqrt{\frac{3}{2}j(j+1)}+{\cal O}\left(\frac{1}{j}\right)\,,
\end{equation}
\begin{equation}
\omega^2=\frac{9/4}{j(j+1)}+{\cal O}\left(\frac{1}{j^3}\right)\,.
\label{omegaj}
\end{equation}
Thus, the eigenvalues (\ref{oscev}) of the effective operator
read to the first leading orders in $j$
\begin{equation}
q_n^{\rm osc}=\frac{4}{3\sqrt{3}}\left(j^3+\frac{3}{2}j^2\right)
-\frac{2}{\sqrt{3}}j^2\left(n+\frac{1}{2}\right)
+{\cal O}\left(j\right)\,.
\end{equation}
In particular, from Eq.~(\ref{omegaj}) we see that the width $1/\sqrt{\omega}$
of the wave functions (\ref{oscwf}) is proportional to $\sqrt{j(j+1)}$.
Moreover, for the largest eigenvalue of the volume operator, one finds
from Eq.~(\ref{defV})
\begin{equation}
\frac{V_0}{(8\pi\gamma\ell_P)^{3/2}}\approx\frac{\sqrt{2}}{3}\sqrt{q^{\rm osc}_0}
=\frac{2^{3/2}}{3^{7/4}}j^{3/2}\left(1-\frac{3}{8}\frac{1}{j}\right)
+{\cal O}\left(\frac{1}{\sqrt{j}}\right)\,.
\end{equation}
Here the leading term $(\propto j^{3/2})$
is exactly the classical volume of a regular tetrahedron whose faces have area
$j$, and the subleading correction is, after a redefinition of the face area, 
identical to the one found in 
Ref.~\cite{Bianchi12} using Bohr-Sommerfeld quantization. Furthermore, our
findings here suggest that the classical volume of a general tetrahedron
with face areas $j_1,j_2,j_3,j_4$ is, to leading order in all $j_i$, given by
\begin{equation}
V_{\rm cl}=\frac{2}{3}\sqrt{\alpha(\bar k)}\,.
\label{Vcl}
\end{equation}

As already discussed in section \ref{tet}, the large eigenvalues $q_n$
have counterparts $q_n^{\prime}=-q_n$ with the the same modulus but negative sign,
and according to Eq.~(\ref{signchange}), the pertaining eigenvectors
can be obtained from the previous ones by changing the sign of any other 
component. Regarding the wave functions (\ref{oscwf}) one could try
to mimic this behavior by attaching an appropriate phase factor,
\begin{eqnarray}
\psi_n^{\prime}(x)=e^{i\pi x}\psi_n(x)\,.
\label{oscev´}
\end{eqnarray}
However, the above functions are clearly not eigenfunctions
of the effective operator (\ref{Qeff}). In fact, an operator having
$\psi_n^{\prime}$ as eigenstates with eigenvalues $(-q_n^{\rm osc})$
can be constructed as follows:
\begin{eqnarray}
Q_{\rm osc}^{\prime} & = & -e^{i\pi x}Q_{\rm osc}e^{-i\pi x}
\label{Qeff´1}\\
 & = & -\bar q\left[1-\left(\frac{1}{2}\left(p-\pi \right)^2
+\frac{\omega^2}{2}x^2\right)\right]\,.\label{Qeff´2}
\end{eqnarray}
This operator is not invariant under a \lq time reversal\rq $p\mapsto -p$
which corresponds to the fact that the eigenfunctions (\ref{oscev´}) cannot
be chosen to be real. Moreover, the operators $Q_{\rm osc}$ and
$Q_{\rm osc}^{\prime}$ are, along with their eigenfunctions,
obviously just related by a U(1) gauge operation, apart from the
global minus sign on the r.h.s of Eqs.~(\ref{Qeff´1}) and (\ref{Qeff´2}).
However, since $Q_{\rm osc}^{\prime}$ is merely a consequence of the rather
phenomenological ansatz (\ref{oscev´}), a more rigorous effective description
of eigenstates with negative eigenvalue is desirable. Work in this direction
could possibly build upon ideas of Ref.~\cite{Aquilanti13} where the
quantity $\pm 2\alpha(k)$ was considered as an effective potential
for states with eigenvalues of both sign.

\section{Recoupling of Angular Momenta}
\label{rec}

There are obviously alternatives to the coupling scheme of angular momenta
we have used so far.
For instance, instead of the previous procedure,
$\vec j_1,\vec j_3$ and $\vec j_2,\vec j_4$ could first be coupled 
to two irreducible representations of dimension $2l+1$ each,
which are then combined  to a total singlet. 
The operator $Q$ is then expressed in a form analogous to Eq.~(\ref{defQ})
with matrix elements $\beta(l)$ given by the r.h.s of Eq.~(\ref{alpha}) 
and obvious interchanges of labels. As seen before, $\beta(l)$ has a
unique maximum at some $l=\bar l$. Thus, putting again
$l=\bar l+y-1/2$, the eigenstates with large eigenvalues will again be
accurately approximated by oscillator wave functions $\psi_n(y;\nu)$
according to Eq.~(\ref{oscwf}) with
\begin{equation}
\nu^2=-\frac{\left(\frac{d^2\beta(l)}{dl^2}\right)_{l=\bar l}}
{\beta(\bar l)}\,,
\label{nu}
\end{equation}
and the corresponding approximate eigenvalues read
\begin{equation}
q_n^{\rm osc}=\bar r\left(1-\nu(n+1/2)\right)
\end{equation}
with $\bar r=2\beta(\bar l)$. 

Since the exact spectrum of $Q$ is of course independent of the coupling scheme
used, this holds as well, to an excellent degree of approximation, for
the approximate eigenvalues, as it is easily checked by numerics. For
instance, for the parameters of Fig.~\ref{fig1} and table ~\ref{table1}
we find $\bar q=13649.6$, $\bar r=13650.4$ and $\omega=0.075206$,
$\nu=0.075198$. Thus, we have, as an again excellent approximation,
\begin{equation}
\omega\approx\nu\,,
\end{equation}
which in particular means that the wave functions 
$\psi_n(x;\omega)$ and $\psi_n(y;\nu)$ can
be taken as identical.

Moreover, since switching to another coupling scheme implies just a change of
basis in the Hilbert space, the above  two gauss-hermitian wave functions 
should be related by a unitary transformation,
\begin{equation}
\eta_n\psi_n(y;\omega)=\int dx\,U(y,x)\psi_n(x;\omega)\,,
\label{trafo1}
\end{equation}
with some phase factor $\eta_n$, $|\eta_n|=1$.
An obvious solution is given by $\eta_n\equiv 1$ and $U(y,x)=\delta(y-x)$, while
another possibility follows from the well-known fact that the Fourier transform
of a gauss-hermitian function is a function of that same type: 
Here one has $\eta_n=(-i)^n$ and
\begin{equation}
U(y,x)=\sqrt{\frac{\omega}{2\pi}}e^{-i\omega yx}\,.
\label{trafo2}
\end{equation}
Thus, up to the scale factor $\omega$ occurring in Eq.~(\ref{trafo2}),
changing from one coupling scheme to another just corresponds to a Fourier
transform of the approximating oscillator wave functions. This observation
is of course strongly reminiscent of switching from real space to momentum
representation in standard quantum mechanics.

On the other hand, treating $k$ and $l$ again as discrete state labels,
the basis states in both coupling schemes are related by
\begin{equation}
|l\rangle=\sum_{k=k_{\rm min}}^{k_{\rm max}}(-1)^{j_2+j_3+k+l}
\sqrt{(2k+1)(2l+1)}
\left\{
\begin{array}{ccc}
 j_1 & j_2 & k \\
 j_4 & j_3 & l 
\end{array}
\right\}|k\rangle\,,
\label{recoup}
\end{equation}
using Wigner $6j$-symbols in the standard convention of prefactors
\cite{Edmonds57}. For the case of all angular momenta being large compared
to unity, Ponzano and Regge \cite{Ponzano68} have devised 
the following asymptotic expression for such quantities
(for more recent developments, see also 
Refs.~\cite{Gurau08,Dupuis09,Aquilanti12}),
\begin{equation}
\left\{
\begin{array}{ccc}
 j_1 & j_2 & j_3 \\
 j_4 & j_5 & j_6 
\end{array}
\right\}\approx\frac{1}{\sqrt{12\pi{\cal V}}}\cos\left(\frac{\pi}{4}+
\sum_{i=1}^6\theta_i(j_i+1/2)\right)\,.
\label{ponzano}
\end{equation}
Here $\cal V$ is the volume of a tetrahedron having edge lengths $(j_i+1/2)$,
$i\in\{1\dots 6\}$ where edges occurring in the same column of the
$6j$-symbol are opposite to each other, i.e. do not have a common vertex,
and $\theta_i$ is the external dihedral angle between faces joining at
edge $j_i$. The cosine occurring in the above equation bears some
similarity to the exponential in the transformation (\ref{trafo2}). Moreover,
under a rescaling of all angular momenta, $j_i\mapsto uj_i, k\mapsto uk,
l\mapsto ul$, $\cal V$
scales obviously as $u^3$, such that the prefactor of $|uk\rangle$ in
Eq.~(\ref{recoup}) is proportional to $1/ \sqrt{u}$, which is the same
scaling behavior as in (\ref{trafo2}). However, we leave it to future
studies to more deeply investigate the possible relationship between
the transformations (\ref{trafo2}) and (\ref{recoup}).

\section{Conclusions and Outlook}
\label{concl}

We have shown that the (square of the) volume operator of a quantum 
tetrahedron is, in the
sector of large egenvalues, accurately desribed by a quantum harmonic
oscillator. This finding is a consequence of the fact that 
(i) the volume operator couples
only neighboring states of its standard basis, and (ii) its matrix elements
show a unique maximum as a function of state labels.
The ingredients of the harmonic oscillator constructed here are 
an appropriate coordinate variable and a momentum operator defined by
the corresponding derivative. These two quantities fulfill the canonical
commutation relation.

We give explicit formulae for the large
eigenvalues of the volume operator in terms of the equidistant 
harmonic oscillator spectrum. It is an interesting speculation whether or 
not these linear excitations of space are related to gravitational waves.
Moreover, in this limit the quantum tetrahedron is naturally described
semiclassically by oscillator coherent states, in contrast to other
approaches where tensor products of 
SU(2) coherent states projected onto the singlet subspace are
used \cite{Livine07,Perez13}.

We have also analyzed the scaling
properties of the oscillator parameters as a function of the size
of the tetrahedron. For a regular tetrahedron we reproduce recent findings
\cite{Bianchi12} on the largest volume eigenvalue and generalize them
to the next smaller eigenvalues. In terms of classical geometry, our approach 
here also suggests an interesting expression given in Eq.~(\ref{Vcl})
for the volume of a general tetrahedron. To further investigate this
conjecture might be, from a more mathemaitcal perspective, a route
for future studies (possibly starting from numerical tests). 
Here we have shown the result only for the very special
case of a regular tetrahedron.
Finally, we have discussed the role of different angular 
momentum coupling schemes.

One might argue that the findings here on the tetrahedral volume 
operator are in fact very general: Expanding a classical system
described by just one pair of canonical variables \cite{Bianchi11b,Bianchi12}
around an extremum will generically lead to an effective harmonic
oscillator. An interesting point here is that this oscillator-like behavior
sets in at already quite moderate lengths of the involved angular
momenta (being about five). Moreover, the quantum number resulting from the
coupling of angular momenta has an immediate interpretation in terms
of the oscillator coordinate.

The present work exclusively deals with tetrahedra, i.e., in the language
of spin networks, 4-valent nodes \cite{Perez13}. An obvious and interesting 
question is how the results found here translate to higher nodes.
Recent work, in a similar spirit as here,  
on the semiclassical properties of pentahedra includes
Refs.~\cite{Haggard13,Coleman-Smith13}.

\section*{Acknowledgements}

I thank Hal Haggard for useful correspondence.

{}

\end{document}